\documentclass[10pt, aps,pre,twocolumn,amsfonts,amssymb,amsmath,showkeys,floatfix]{revtex4-2}
\usepackage{graphicx}           
\usepackage[colorlinks,linkcolor=blue,citecolor=blue,breaklinks=true]{hyperref}
\usepackage[caption=false,subrefformat=parens,labelformat=parens]{subfig}
\usepackage{bm}          
\usepackage{booktabs}
\usepackage{tabularx}
\usepackage{url}                
\usepackage{color}
\usepackage{mathtools}
\usepackage{accsupp}            
\usepackage{listings}  
\usepackage{algorithm}
\usepackage{algpseudocode}
\usepackage{threeparttable} 
\usepackage{easyReview}

\renewcommand*{\vec}[1]{\bm{\mathrm{#1}}}  
\newcommand*{\vechat}[1]{\vec{\hat{#1}}}   
\newcommand{\tc}{T_\mathrm{c}} 

\newcommand{\subfigref}[2]{\hyperref[#1]{\ref*{#1}(#2)}}
\newcommand{\spacedsubfigref}[2]{\hyperref[#1]{\ref*{#1}~(#2)}}
\begin{document}

\title{Critical Behavior Analysis of Pure Dipolar Triangular Lattice via Equilibrium and Non-Equilibrium Monte Carlo Simulations}

\author{S. Ismailzadeh}
\author{M. D. Niry}
\thanks{Author to whom correspondence should be addressed.
	Electronic address: \url{m.d.niry@iasbs.ac.ir}; 
	URL: \url{http://www.iasbs.ac.ir/~m.d.niry/}}
\affiliation{Department of Physics, Institute for Advanced Studies in Basic Sciences (IASBS), Zanjan 45137-66731, Iran}

\date{\today}

\begin{abstract}
Magnetic thin films and 2D arrays of magnetic nanoparticles exhibit unique physical properties that make them valuable for a wide range of technological applications. In such systems, dipolar interactions play a crucial role in determining their physical behavior. However, due to the anisotropic and long-range nature of dipolar interactions, conventional Monte Carlo (MC) methods face challenges in  investigating these systems near criticality. In this study, we examine the critical behavior of a triangular lattice of XY dipoles using the optimized Tomita MC algorithm tailored for dipolar interactions. We employ two independent computational approaches to estimate the critical temperature and exponents: equilibrium MC simulations with histogram reweighting and the non-equilibrium relaxation method. Notably, both approaches demonstrate that this XY dipolar system might be in a new universality class very close to the 2D Ising universality class.
\end{abstract}
\maketitle

\section{\label{sec:int}Introduction}
Two-dimensional arrays of magnetic nanoparticles and magnetic thin films have attracted considerable attention due to their unique properties and diverse applications in spintronics \cite{karmakar2011}, data storage \cite{sun2004}, and biomedicine \cite{peixoto2020}. In nanoparticle assemblies, the large magnetic moments of the particles lead to significant dipolar interactions \cite{kechrakos2008}. These interactions are particularly important in these systems, as they cause the system's magnetic properties to depend on the nanoparticles' geometric arrangement. For example, simple point dipoles on a triangular lattice align parallel in the ground state, whereas a square array forms anti-parallel magnetic stripes \cite{tomita2009, russier2001, debell1997}. Therefore, understanding emergent behaviors in these systems, particularly near a phase transition, requires an in-depth investigation of dipolar interactions and their effects on system behavior. 
 
A crucial point concerning such 2D XY systems is the Mermin-Wagner theorem, which forbids long-range magnetic order for systems with continuous symmetry and short-range interactions \cite{mermin1966}. While the 2D dipolar systems can be considered short-range by some definitions i.e., its spatial integral converges \cite{campa2014}, it is long-range enough to circumvent this theorem \cite{bouchet2010}. The interaction fundamentally alters the low-energy spin-wave excitations, making their dispersion stiffer than in typical short-range systems. This stiffening suppresses the long-wavelength fluctuations that would otherwise destroy the order, thus allowing for a stable ordered phase at a finite temperature \cite{bouchet2010}. The stabilization of long-range order has been shown in various contexts: for isotropic XY spins with $1/r^\alpha$  interactions where $2 < \alpha < 4$  \cite{kunz1976}, and in 2D Heisenberg systems where a weak dipolar interaction is added to a strong exchange interaction \cite{maleev1976}. Specifically for pure 2D XY dipolar lattices, the existence of this finite-temperature phase transition was demonstrated theoretically \cite{malozovsky1991}. Subsequent Monte Carlo simulations have confirmed this long-range order, revealing a continuous phase transition on square and triangular dipolar lattices \cite{tomita2009, debell1997, rastelli2002}.

Critical phenomena, which describe the behavior of systems near continuous phase transitions, are particularly important because they reveal universal behaviors across diverse systems \cite{goldenfeld1992, kardar2007}. However, the analysis of critical phenomena with Monte Carlo simulations poses unique challenges due to critical slowing down \cite{newman1999, janke2008}. For systems dominated by dipolar interactions, these challenges become even more pronounced because of the long-range and anisotropic nature of the dipole-dipole interactions. The computational complexity of simulating these systems using traditional Metropolis algorithm, becomes $\mathcal{O}(N^2)$  due to the long-range nature of dipolar interactions.

While MC methods exist to alleviate the problems of critical slowing down and $\mathcal{O}(N^2)$ complexity in long-range interacting systems \cite{luijten1995, fukui2009, kapfer2016}, they are not efficient in simulating dipolar systems due to anisotropy of dipolar interactions. Consequently, suitable methods for simulating dipolar interactions near the critical point are very limited. These challenges have constrained research on the critical properties of dipolar systems.

Recent advancements have introduced MC methods with $\mathcal{O}(N)$ complexity applicable for dipolar interactions \cite{michel2019, sasaki2008, tomita2009, muller2023}. However, these methods were not originally tailored to optimize performance in dipolar systems. In our recent work, we systematically compared and adjusted three advanced algorithms—clock, SCO, and Tomita—to adapt them to such systems. Our adjustments led to significant improvements in computational efficiency, enabling higher-precision simulations of critical properties in dipolar systems \cite{ismailzadeh}.

In this study, we employed the optimized Tomita MC algorithm for dipolar interactions to investigate the critical behavior of a triangular lattice of XY dipoles using both equilibrium and non-equilibrium approaches.
In the equilibrium method, we simulated the system near the critical temperature and applied histogram reweighting combined with finite-size scaling analysis to determine the critical temperature and exponents \cite{newman1999, chen1993}.
For the non-equilibrium approach, we employed the non-equilibrium relaxation (NER) method. This technique analyzes the relaxation dynamics of the system as it evolves from a fully magnetized state at temperatures close to the critical point. Extracting the critical temperature and exponents from the relaxation dynamics is straightforward \cite{ozeki2007}.

\section{Methods}
In this section, we first introduce the model and its related thermodynamic quantities. Then, we explain the equilibrium method, and finally, we discuss the non-equilibrium relaxation technique.

\subsection{Model, thermodynamics and Monte Carlo simulations}
The Hamiltonian for a two-dimensional lattice of spins with dipolar interactions is given by
\begin{equation}
\mathcal{H} (\{\vec{S}_i\}) = - \frac{1}{2} \frac{\mu_0}{4\pi} \sum_{\substack{i, j \\ i \ne j}}\frac{ 3(\vec{S}_i\cdot\vechat{r}_{ij})(\vec{S}_j\cdot\vechat{r}_{ij}) - \vec{S}_i\cdot\vec{S}_j}{|\vec{r}_{ij}|^3}. \label{eq:hamiltonian}
\end{equation}
where \(\vec{S}_i\) denotes the spin vector located at site \(i\), and the summation accounts for the dipolar interactions among all spin pairs. The vector \(\vec{r}_{ij} = |\vec{r}_{ij}| \, \vechat{r}_{ij}\) specifies the relative position of spin \(j\) with respect to spin \(i\). The factor $1/2$ is due to double counting and the constant \(\mu_0\) is the vacuum permeability.

To investigate the critical behavior of a system, we are particularly intersted in the themodynamic quantities.	for a magnetic system the order parameter is defined as the average magnetization per spin,
\begin{equation}
m= \frac{1}{N} \left\langle M \right\rangle,
\end{equation}
In the triangular lattice of XY dipoles, the spins align within the lattice plane in the ground state. Therefore, we define
\begin{equation}\label{key}
M = \left| \sum_{i=1}^{N}\vec{S}_i \right| .
\end{equation}

For a thermodynamic system, key response functions can be calculated from equilibrium fluctuations. Two important response functions in a magnetic system are the magnetic susceptibility $\chi$, related to fluctuations in magnetization $M$, and the heat capacity $c$, related to fluctuations in energy $E$ \cite{newman1999,janke2008}:
\begin{align}
\chi&=\frac{\left\langle M^{2}\right\rangle-\langle M\rangle^{2}}{N k_{\mathrm{B}} T}, \\
c&=\frac{\left\langle E^{2}\right\rangle-\langle E\rangle^{2}}{N k_{\mathrm{B}} T^{2}},
\end{align}

In a system Near the critical point, the thermodynamic quantities exhibit power-law behavior following
\begin{subequations} \label{eq:scaling_relations}
\begin{align}
c & \sim |\varepsilon|^{-\alpha},  \label{eq:scaling_c} \\
m & \sim |\varepsilon|^{\beta} \quad \text{for } T < T_\mathrm{c}, \label{eq:scaling_m} \\
\chi & \sim |\varepsilon|^{-\gamma}, \label{eq:scaling_chi} \\
\xi & \sim |\varepsilon|^{-\nu}, \label{eq:scaling_xi}
\end{align}
\end{subequations}
where  $\varepsilon= (T - T_\mathrm{c})/T_\mathrm{c}$  is the reduced temperature, $\xi$  is the correlation length and \( \alpha \), \( \beta \), \( \gamma \), and \( \nu \) are the critical exponents \cite{kardar2007}.

A remarkable feature of these exponents is their universality which states that seemingly different thermodynamic systems exhibit the same exponents if they share certain fundamental properties. Consequently, systems can be grouped into a relatively small number of universality classes. The class to which a system belongs is determined by general properties like the spatial dimension $d$, the symmetries of the order parameter, and the range of the microscopic interactions (e.g., short-range vs. long-range), rather than by the specific microscopic details of the system \cite{kardar2007,goldenfeld1992}.

In many systems,
the critical exponents obey the scaling and hyperscaling relations
\begin{equation} \label{eq:scaling_hyper}
2-\alpha=d \nu=2 \beta+\gamma ,
\end{equation}
where $d$  is the spatial dimension. Therefore, only two of these critical exponents are independent \cite{kardar2007, goldenfeld1992}.

The spins in this study are considered to be planar rotors, i.e. $O(2)$  spins. All quantities are expressed in reduced units, where the spins magnitude and the lattice constant normalized to 1, also we set $\mu_0=8\pi k_\mathrm{B}$. The shape of simulation cell is chosen parallelogram with sides $L$. Periodic boundary conditions are applied, and long-range interactions are accounted for using the Ewald summation method \cite{ewald1921,mazars2011}.

For MC simulations, the optimized version of the Tomita method tailored for dipolar interactions is employed to ensure fast simulation with $\mathcal{O}(N)$ complexity \cite{tomita2009, ismailzadeh}. One MC step (MCS) consists of an attempt to assign new directions to all spins in the lattice. 
Two types of MCSs were used: random steps, in which the new spin direction is chosen randomly, and over-relaxation steps, in which the new direction of a spin is the reflection of its old direction relative to the effective magnetic field on the spin. The over-relaxation steps are employed to accelerate the dynamics  \cite{creutz1987}. Notably, the over-relaxation steps in the Tomita method are stochastic rather than rejection-free, so they do not reduce the dynamic exponent from $z \approx 2$ (see Ref.~\citenum{brown1987}). Nevertheless, they remain effective for reducing autocorrelation time, and the algorithm's overall efficiency is largely independent of the chosen ratio of over-relaxation to random moves \cite{ismailzadeh}. For details of optimized Tomita method for dipolar interactions see Refs.~\citenum{tomita2009} and \citenum{ismailzadeh}. In this study errors in quantities are estimated using the bootstrap method \cite{efron1994}.

\subsection{Equilibrium method}

\subsubsection{Histogram reweighing}
A state \(\phi\) at inverse temperature \(\beta = 1/k_B T\) appears in the configuration space with the Boltzmann probability \(p_{\beta}(\phi) \propto e^{-\beta E_\phi}\) \cite{blundell2009}. The probability of observing this state at  a different inverse temperature \(\beta'\) is related to its probability at \(\beta\) through the following relation \cite{janke2008}:
\begin{equation}\label{boltzmanprob}
p_{\beta^{\prime}}(\phi) \propto e^{-\beta^{\prime} E_\phi} \propto e^{-\left(\beta^{\prime}-\beta\right) E_\phi} p_\beta(\phi).
\end{equation}

The expected value of a quantity at inverse temperature \(\beta\) is given by
\begin{equation}\label{key}
\langle Q\rangle_{\beta} = \frac{\int d \phi \; Q(\phi) p_{\beta}(\phi)}{\int d \phi \; p_{\beta}(\phi)}.
\end{equation}
Using Eq. \eqref{boltzmanprob}, the expectation value of a quantity at a different inverse temperature \(\beta'\) can be expressed in terms of the expectation value at \(\beta\) as
\begin{equation}\label{ReweightingGeneral}
\langle Q\rangle_{\beta^{\prime}}=\frac{\left\langle Q e^{-\left(\beta^{\prime}-\beta\right) E}\right\rangle_\beta}{\left\langle e^{-\left(\beta^{\prime}-\beta\right) E}\right\rangle_\beta}.
\end{equation}

In MC methods, importance sampling is usually employed, meaning configurations are generated according to the Boltzmann distribution at \(\beta\). Because these configurations are already sampled with probabilities \(p_\beta(\phi)\), the ensemble averages in Eq. \eqref{ReweightingGeneral} can be directly estimated using the generated configurations \cite{newman1999}. Consequently, Eq. \eqref{ReweightingGeneral} simplifies for MC simulations to a weighted sum over configurations sampled at \(\beta\):
\begin{equation}\label{key}
\langle Q\rangle_{\beta^{\prime}}=\frac{\sum_i Q_i e^{-\left(\beta^{\prime}-\beta\right) E_i}}{\sum_i e^{-\left(\beta^{\prime}-\beta\right) E_i}}
\end{equation}
where \(Q_i\) and \(E_i\) are the values of \(Q\) and \(E\) for the \(i\)th configuration sampled at \(\beta\) \cite{newman1999,janke2008,ferrenberg1988}.  

This reweighting relation implies that a single simulation at \(\beta\) can, in principle, estimate \(\langle Q \rangle_{\beta'}\) at any other temperature by re-scaling the contributions of each configuration with the factor \(e^{-(\beta'-\beta)E_i}\). However, in practice, the finite number of configurations sampled during the MC simulation limits the reliability of this method. The exponential weights \(e^{-(\beta'-\beta)E_i}\) can vary drastically for \(\beta'\) far from \(\beta\), causing a few configurations to dominate the sum and introducing large statistical errors. Thus, reweighting is only effective for temperatures \(\beta'\) close to the original simulation temperature \cite{ferrenberg1988, newman1999}.

\subsubsection{Finite-size scaling analysis}
Because numerical simulations use finite systems, the correlation length close to the critical temperature, though very large, is always limited. This means that any divergences in calculated quantities will be smoothed out and shifted. This phenomenon is explained by finite-size scaling (FSS) theory. At its core, FSS theory states that near the critical temperature $\tc$, the linear size $L$ of the system effectively replaces the correlation length $\xi$ as the dominant length scale governing the system's behavior \cite{newman1999, fisher1972} . Consequently, the scaling laws in Eqs. \eqref{eq:scaling_relations} are modified according to the FSS ansatz as 
\begin{subequations} \label{eq:fss_relations}
\begin{align}
& m(T, L) \approx L^{-\beta / \nu} \mathcal{M}\left(\varepsilon L^{1 / \nu}\right) \label{eq:fss_m}, \\
& \chi(T, L) \approx L^{\gamma / \nu} \mathcal{X}\left(\varepsilon L^{1 / \nu}\right)  \label{eq:fss_chi}, \\
& c(T, L) \approx c_{\infty}(\varepsilon)+L^{\alpha / \nu} \mathcal{C}\left(\varepsilon L^{1 / \nu}\right), \label{eq:fss_c}
\end{align}
\end{subequations}
where  $\mathcal{M}(x)$ , $\mathcal{X}(x)$ , and $\mathcal{C}(x)$ are scaling functions.

Multiple approaches exist for determining the location of the critical point. One approach involves using the Binder parameter \cite{binder1981}:
\begin{equation}
U_{4}=1-\frac{\left\langle M^{4}\right\rangle}{3\left\langle M^{2}\right\rangle^{2}}.
\end{equation}
It can be shown that the value of the Binder parameter at the critical point is independent of the system size \cite{binder1981}. Consequently, the critical point can be determined as the intersection point of the Binder parameter curves for different system sizes.

Another approach to locate the critical point is to use the following set of quantities to simultaneously determine the critical exponent $\nu$ and the critical point, as introduced by Chen et al. \cite{chen1993}:
\begin{equation}\label{key}
\begin{aligned}
& V_{1} \equiv 4\left[M^{3}\right]-3\left[M^{4}\right], \\
& V_{2} \equiv 2\left[M^{2}\right]-\left[M^{4}\right], \\
& V_{3} \equiv 3\left[M^{2}\right]-2\left[M^{3}\right], \\
& V_{4} \equiv\left(4[M]-\left[M^{4}\right]\right) / 3, \\
& V_{5} \equiv\left(3[M]-\left[M^{3}\right]\right) / 2, \\
& V_{6} \equiv 2[M]-\left[M^{2}\right],
\end{aligned}
\end{equation}
where $\left[M^{n}\right]$ represents the logarithm of the derivative of the $n$th moment of the magnetization $M$:
\begin{equation}
\left[M^{n}\right] \equiv \ln \frac{\partial\left\langle M^{n}\right\rangle}{\partial T}.
\end{equation}

The quantities \(V_1\) to \(V_6\) near the critical point behave  as follows \cite{chen1993}:  
\begin{equation}\label{eq:vjscaling}
V_j(T, L) \approx \frac{1}{\nu} \ln L + \mathcal{V}_j\left(\varepsilon L^{1/\nu}\right).
\end{equation}
where $\{\mathcal{V}_j\}$ are scaling functions. At \(\tc\) (\(\varepsilon = 0\)), the scaling function \(\mathcal{V}_j(0)\) reduces to a constant, so fitting \(V_j\) versus \(\ln L\) yields a slope \(1/\nu\) independent of \(j\). Away from \(\tc\), the \(L\)-dependence of \(\mathcal{V}_j\left(\varepsilon L^{1/\nu}\right)\), causes deviations in the slope estimates.  

To determine \(\tc\) and \(\nu\), multiple temperatures are scanned. The critical point is identified as the temperature where all \(1/\nu\) estimates from \(V_1\) to \(V_6\) converge within statistical uncertainties. The uncertainty in \(\tc\) corresponds to the temperature range over which convergence persists. The exponent \(\nu\) and its error are derived from the mean and standard deviation of the converged \(1/\nu\) values.

Another method for finding the critical point involves using the location of the extremum of the magnetic susceptibility and also the following quantities introduced by Chen et al. \cite{chen1993}:
\begin{subequations}
\begin{align}\label{key}
D_{K_{2}} &\equiv \frac{\partial\left\langle(M-\langle M\rangle)^{2}\right\rangle}{\partial T},\\
D_{K_{3}} &\equiv \frac{\partial\left\langle(M-\langle M\rangle)^{3}\right\rangle}{\partial T},\\
P_{3} &\equiv \frac{\left\langle(M-\langle M\rangle)^{3}\right\rangle}{\langle M\rangle\left\langle(M-\langle M\rangle)^{2}\right\rangle},\\
P_{4} &\equiv \frac{3\left\langle(M-\langle M\rangle)^{2}\right\rangle^{2}-\left\langle(M-\langle M\rangle)^{4}\right\rangle}{3\left\langle M^{2}\right\rangle^{2}}.
\end{align}
\end{subequations}
According to the FSS hypothesis, the location of the extrema of these quantities obeys \cite{chen1993}:
\begin{equation} \label{eq:fss_extermum}
\tc(L) \approx \tc + a_q L^{-1 / \nu},
\end{equation}
where \(\tc\) is the finite-size critical temperature (i.e., the location of the extremum), and \(a_q\) is a quantity-dependent constant. Once the value of the critical exponent \(\nu\) is known, the critical temperature \(\tc\) can be estimated by fitting \(\tc(L)\) as a function of \(L^{-1 / \nu}\).

\subsection{Non-equilibrium relaxation method}
In the non-equilibrium relaxation method, the relaxation dynamics of the system from an initial magnetization at a temperature $T$ is considered \cite{ozeki2007, albano2011}. The behavior of the relaxation process varies depending on whether the system is above, below, or at the critical point.
In particular, initializing the system in its ground state simplifies the dynamics.  In this case, the asymptotic behavior of the magnetization per spin becomes \cite{ozeki2007}:
\begin{equation}\label{eq:NER_simple}
m(t) = \frac{1}{N} \langle {M} \rangle_t \sim \begin{cases}\exp (-t / \tau) & T>T_{\mathrm{c}}\\ t^{-\lambda_{\mathrm{m}}} & T=T_{\mathrm{c}} \\
m_{\mathrm{eq}} + (1-m_{\mathrm{eq}})\exp (-t / \tau)  & T<T_{\mathrm{c}}\end{cases}
\end{equation}
Here, time is measured in terms of MCSs, $\langle \cdots \rangle_t$  represents the ensemble average at time $t$, and $m(t)$  denotes  the magnetization per spin at time $t$. 
In a logarithmic plot of magnetization versus time, the distinct behaviors of the processes in above, below, or at the critical point are reflected in the concavity of the curves. Specifically, the graphs are downward-concave for $T > T_{\mathrm{c}}$, upward-concave for $T < T_{\mathrm{c}}$, and linear at $T = T_{\mathrm{c}}$. This characteristic feature can be used to identify the critical point.

The influence of finite-size effects in an NER process is governed by the growth of the time-dependent dynamical correlation length, $\xi(t)$. Our simulations start from the fully ordered ground state where $\xi(t=0)=0$. As the system evolves at a finite temperature, thermal fluctuations cause correlations to build up, and $\xi(t)$ grows. For as long as this correlation length is much smaller than the system size ($\xi(t) \ll L$), the system behaves as if it were in the thermodynamic limit, causing the relaxation curves for different system sizes to overlap. 

Finite-size effects become significant when $\xi(t)$ becomes comparable to $L$. At this point, the system feels its finite boundaries, and its relaxation deviates from the infinite-system behavior, beginning to relax towards its size-dependent equilibrium magnetization, $m_{\mathrm{eq}}(L)$. A smaller system reaches this crossover condition ($\xi(t) \sim L$) at an earlier time. Therefore, the time interval during which a system of size $L$ remains effectively in the thermodynamic limit can be determined by comparing its relaxation behavior to that of a larger system.

The behaviors summarized in Eq. \eqref{eq:NER_simple} are derived from a more general scaling theory. Near the critical temperature, the magnetization, starting from a fully magnetized state, is described by the relation \cite{albano2011,zheng1998}:
\begin{equation} \label{eq:A1}
m(t,T,L) = b^{-\beta/\nu}\mathcal{M}\left(b^{-z}t, b^{1/\nu}\varepsilon, L/b\right),
\end{equation}
where $\varepsilon=\left(T-\tc\right)/\tc$ is the reduced temperature, $b$ is an arbitrary spatial rescaling factor, $\beta$ and $\nu$ are static critical exponents, $z$ is the dynamical exponent, and $\mathcal{M}$ is a universal scaling function.

By choosing the rescaling factor $b$ to be the time-dependent correlation length, $\xi(t) \sim t^{1/z}$, the relation simplifies to \cite{albano2011}:
\begin{equation} \label{eq:A2}
m(t,T,L) = t^{-\lambda_m} \widetilde{\mathcal{M}}\left(\left(\frac{\xi(t)}{\xi_{\text{eq}}(T)}\right)^{1/\nu}, \frac{L}{\xi(t)}\right),
\end{equation}
where $\widetilde{\mathcal{M}}$ is another scaling function, $\xi_{\text{eq}}(T) \sim |\varepsilon|^{-\nu}$ is the equilibrium correlation length, and the exponent $\lambda_m$ is given by
\begin{equation}
\lambda_{\mathrm{m}}=\frac{\beta}{z \nu} \label{eq:lambda_m} . 
\end{equation}

Equation \eqref{eq:A2} predicts a power-law decay, $m(t) \sim t^{-\lambda_m}$, when the scaling function $\widetilde{\mathcal{M}}$ is approximately constant. A true phase transition occurs only in the limit $(L, \; \xi_{\text{eq}} )\to \infty$, so the scaling function is constant only when its first argument is very small and its second is very large. This imposes two conditions on the dynamics: first ${\xi}(t) \ll L$, which implies the system must be effectively in the thermodynamic limit, and second ${\xi}(t) \ll {\xi}_{\text{eq}}(T)$, which states the system must be far from its final equilibrium state \cite{albano2011}.

Away from the critical point, $\xi_{\text{eq}}(T)$ is finite, and the power-law decay is eventually cut off when $\xi(t)$ approaches $\xi_{\text{eq}}(T)$. At the critical point ($T = \tc$), however, $\xi_{\text{eq}}(\tc) \to \infty$, so the power-law behavior is limited only by the finite system size $L$. The NER method uses this to identify $\tc$ as the unique temperature that exhibits a sustained power-law decay. 

The following quantities are introduced by Ito et al. \cite{ito2000} to conveniently determine the critical exponents:
\begin{subequations}
\begin{align}
f_{\mathrm{mm}}(t) & =N\left(\frac{\langle{M}^{2}\rangle_{t}}{\langle{M}\rangle_{t}^{2}}-1\right), \label{eq:fmm} \\
f_\mathrm{m e}(t) & =N\left(\frac{\langle{M} {E}\rangle_{t}}{\langle{M}\rangle_{t}\langle{E}\rangle_{t}}-1\right). \label{eq:fme}
\end{align}
\end{subequations}
Similar to the magnetization, the NER process of these functions exhibit power-law behavior at the critical temperature:
\begin{subequations}
\begin{align}
f_\mathrm{m m}(t) & \sim t^{\lambda_\mathrm{m m}}, \label{eq:scaling_fmm} \\
f_\mathrm{m e}(t) & \sim t^{\lambda_\mathrm{m e}}, \label{eq:scaling_fme}
\end{align}
\end{subequations}
where, the exponents $\lambda_\mathrm{m m}$ and $\lambda_\mathrm{m e}$ are equal to \cite{ozeki2007}:
\begin{subequations}
\begin{align}
\lambda_\mathrm{m m} & =\frac{2 \beta+\gamma}{z v}=\frac{d}{z}, \label{eq:lambda_mm} \\
\lambda_\mathrm{m e} & =\frac{1}{z v}. \label{eq:lambda_me}
\end{align}
\end{subequations}
Here, $d$ is the dimension of the system. 

From the system of Eqs. \eqref{eq:lambda_m}, \eqref{eq:lambda_mm} and \eqref{eq:lambda_me}, the dynamical exponent $z$ and the critical exponents $\beta$ and $\nu$ can be derived as:
\begin{subequations} \label{eq:exponent_by_lambda}
\begin{align}
z & =\frac{d}{\lambda_\mathrm{m m}},  \\
\nu & =\frac{\lambda_\mathrm{m m}}{d \cdot \lambda_\mathrm{m e}},\\
\beta & =\frac{\lambda_\mathrm{m}}{\lambda_{m e}}.
\end{align}
\end{subequations}
To determine the critical exponents, the magnetization and the functions \(f_{\mathrm{mm}}(t)\) and \(f_{\mathrm{me}}(t)\) at the critical point are first plotted on a log-log scale. The slopes of these functions are measured to obtain \(\lambda_\mathrm{m}\), \(\lambda_\mathrm{m m}\), and \(\lambda_\mathrm{m e}\), which are then substituted into the Eqs. \eqref{eq:exponent_by_lambda} to calculate the critical exponents.

\section{Results}

\subsection{Equilibrium results}
First, simulations were conducted at several temperatures with lattice sizes ranging from $L=16$, 24, 32, 48 and $64$. Using Binder's parameter, the critical point was estimated to be at $T_\mathrm{c} = 0.845$.
Next, another set of simulations was performed at this approximate critical temperature for lattice sizes $L=16$, 24, 32, 48, 64, 96 and 112.  The histogram reweighting method was employed to calculate thermodynamic quantities at nearby temperatures.
Subsequently, using parameters $\{V_i\}$  and Binder parameter $U_4$, a refined estimate for the critical point was obtained at $T_\mathrm{c} = 0.849$. 

The final set of simulations was conducted  at this updated critical temperature for lattice sizes $L=16$, 24, 28, 32, 36, 40, 48, 56, 64, 72, 80, 88, 96, 112, 128, 144, 160, 196, 226 and 256.  The histogram reweighting method was again used to calculate thermodynamic quantities at neighboring temperatures.
In each simulation of the final set, one random Tomita MCS was followed by ten overrelaxation Tomita MCS. Each simulation started with an equilibration phase of $10^6$ MCS, after which data were collected over $10^7$ MCS. Each configuration was run 12 times to ensure statistical reliability. All the results presented in this section correspond to the final set of simulations.

The number of independent measurements in the simulations can be estimated using
$n_\mathrm{eff} \simeq {t_\mathrm{max}}/{2\tau},$
where $t_\mathrm{max}$ is the length of the MC simulation and $\tau$ is the autocorrelation time \cite{newman1999}.
Using this relation, the total number of independent measurements across all runs is estimated to range from \(10^6\) for $L=16$  to \(6\times10^3\) for  $L=256$, as the autocorrelation time increases with system size.

In estimating the critical point using the final simulation series, we first utilized the parameters $\{V_i\}$. For this purpose,  we calculated the inverse exponent \(1/\nu\) by determining the slope of the parameters \(\{V_i\}\) with respect to \(\ln L\), following Eq. \eqref{eq:vjscaling}.  At the critical point, the calculated $1/\nu$ parameters are identical for each $V_j$, while they differ at other temperatures.

\begin{figure}[tbp]
\centering\includegraphics[width=\columnwidth]{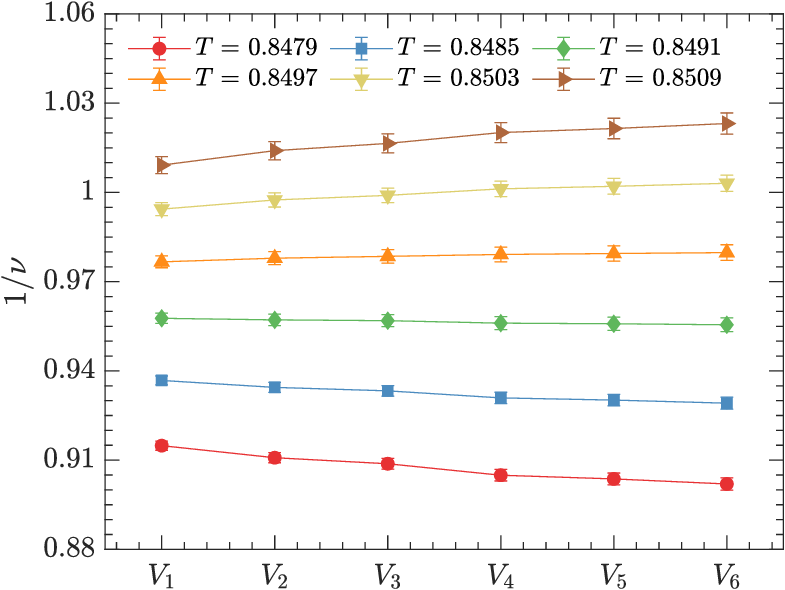}
\caption{The values of \(1/\nu\) determined using the set of parameters \(\{V_i\}\) at varying temperatures \(T = 0.8479\) to \(0.8509\). These exponents are derived from linear fits to \(V_i\) as a function of system size in semi-log plots, following Eq. \eqref{eq:vjscaling}. The lines are drawn only to guide the eye.}
\label{fig:VivsInvNu}
\end{figure}

The \( 1/\nu \) values obtained  using the parameters  \( V_1 \) to \( V_6 \) at various temperatures are shown in Fig. \ref{fig:VivsInvNu}. For the temperature interval \( [0.8489, \,  0.8497] \), the exponents \( 1/\nu \) calculated using each \( V_i \) are equal within the error range. This can be seen in the figure where data points are aligned along a horizontal line. This temperature interval is therefore identified as the uncertainty range for the critical temperature. Consequently, the critical point is estimated as \( \tc = 0.8493(4) \), and the inverse critical exponent \( 1/\nu = 0.96(2) \), which corresponds to \( \nu = 1.04(2) \).

\begin{figure}[tbp]
\centering\includegraphics[width=\columnwidth]{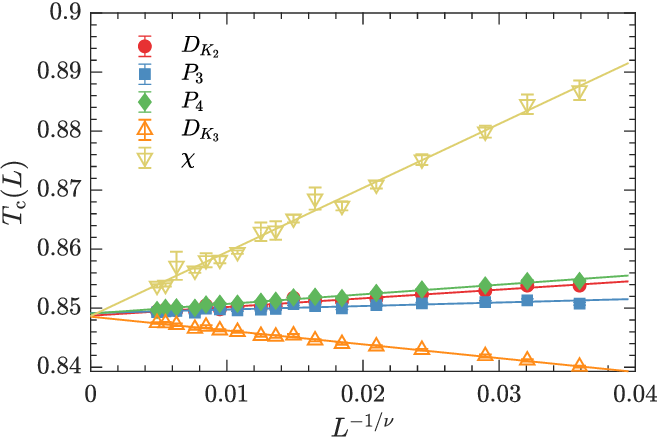}
\caption{Size-dependent behavior for the location of extrema \(\tc(L)\) of various thermodynamic quantities. The straight lines represent the linear fits to the data, assuming a scaling exponent of $1/\nu=0.96$.  Quantities whose extrema are closer to the critical temperature are indicated with solid markers.}
\label{fig:extermaVsL}
\end{figure}

Using the estimated critical exponent $\nu$, we apply Eq.~\eqref{eq:fss_extermum} to achieve a more accurate estimate of the critical point. To do this, we first plot the extrema locations of various quantities as a function of $L^{-1/\nu}$ for system sizes $L = 32$ to $256$, as shown in Fig. \ref{fig:extermaVsL}. We then perform a linear extrapolation of these extrema locations toward the thermodynamic limit ($L \to \infty$). The y-intercept of these lines corresponds to the critical point. Using this method, we estimate the critical temperature to be $ T_\mathrm{c} = 0.8488(3) $. The coefficients of determination ($ R^2 $) for the fits were generally above 0.9, except for the parameter $ P_3 $, which had an $ R^2 $ of 0.7.

By combining the new estimate of the critical temperature with the previous one from the set of $\{V_i\}$, we derive a refined critical temperature of $ T_\mathrm{c} = 0.8491(4) $. 
The estimate for \(1/\nu\) based on Eq. \eqref{eq:vjscaling} depended on the accuracy of our critical point estimation. With the new estimate of \(T_\mathrm{c}\), the inverse critical exponent is updated to \( 1/\nu = 0.955(15) \), yielding \( \nu = 1.05(2) \).  

\begin{figure}[tbp]
\centering
\includegraphics[width=\columnwidth]{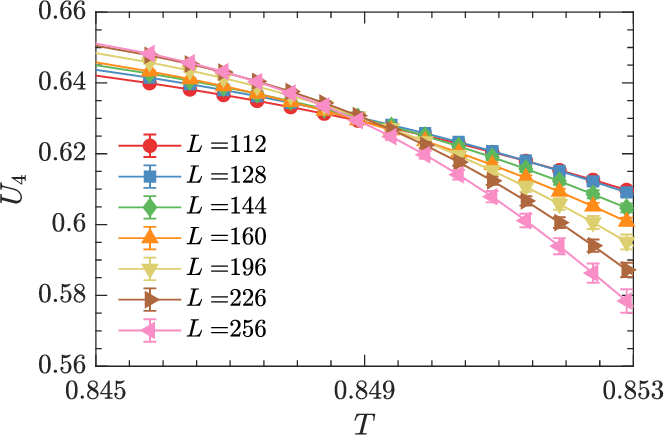}
\caption{Binder parameter \( U_4 \) versus temperature \( T \) for different system sizes \( L \).
The curves are drawn to guide human eye.}
\label{fig:binder}
\end{figure}

The Binder parameter for system sizes $ L = 112$ to 256 is plotted in Fig. \ref{fig:binder}. The intersection point of the curves indicates the critical point, located near $ T_\mathrm{c} \simeq 0.849 $, and aligns well with the other estimates.  

From Eqs. \eqref{eq:fss_m} and \eqref{eq:fss_chi}, we determine the ratios \( {\beta}/{\nu} \) and \( {\gamma}/{\nu} \) using the slopes of magnetization and susceptibility at the critical point as a function of system size in the log-log plot:  
\begin{equation}\label{key}  
\frac{\beta}{\nu} = 0.153(6), \quad \frac{\gamma}{\nu} = 1.70(2).  
\end{equation}  

\begin{figure}[tbp]
\centering
\includegraphics[width=\columnwidth]{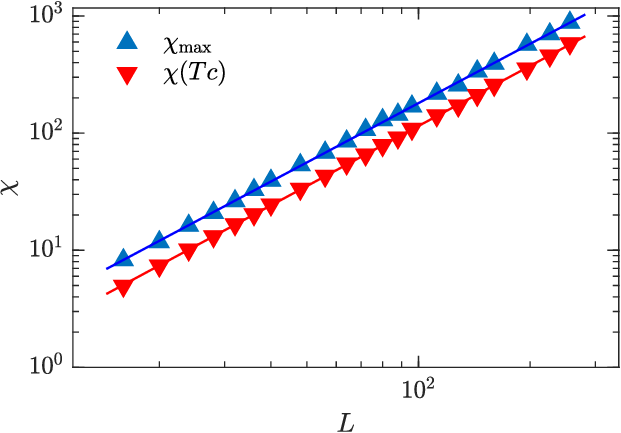}
\caption{Maximum susceptibility and susceptibility at estimated $T_\mathrm{c}=0.8491$ as a function of system sizes $L$. The errors are smaller than the symbol sizes, and the lines represent best fits to the data.}
\label{fig:chi_two_method}
\end{figure}

Another method for estimating the ratio \( {\gamma}/{\nu} \) involves calculating the slope of the maximum susceptibility against system size in a log-log plot. This relies on the fact that the scaling function $\mathcal{X}$ in Eq. \eqref{eq:fss_chi} is the same for all system sizes, so its maximum value is independent of \( L \) \cite{newman1999}. Using this method, we determine  
\begin{equation}\label{key}  
\frac{\gamma}{\nu} = 1.68(1).  
\end{equation}  
The error from this method is smaller because it is not affected by the uncertainty in the critical point.
The two methods to compute ${\gamma}/{\nu}$ are illustrated in Fig. \ref{fig:chi_two_method}. As shown, the slopes of the fittings are roughly equal. By averaging the two estimates, we obtain:  
\begin{equation}\label{key}
\frac{\gamma}{\nu} = 1.69(1).
\end{equation}

We calculated additional estimates for \( \nu \) using the logarithmic derivatives of the magnetization. Specifically, we analyzed the slopes of $\partial_\beta \log \langle m \rangle$ and $\partial_\beta \log \langle m^2 \rangle$ versus system size in a log-log plot at the critical point \cite{janke2008}:  
\begin{equation}\label{key}  
\frac{1}{\nu_1} = 0.961(17), \quad \frac{1}{\nu_2} = 0.967(15).  
\end{equation}  
These values are consistent with the previous estimate of \( 1/\nu \), further validating the results.  

All regression analyses for estimating the critical exponents yielded  \( R^2 \) values above 0.99, demonstrating a high goodness of fit.

From this analysis, the final results for the critical parameters are as follows:
\begin{equation}\label{eq:final_equilibrium}
\begin{aligned}
    \tc   & = 0.8491(4), \quad  & \nu    & = 1.05(2),\\
    \beta & = 0.160(7),  \quad  & \gamma & = 1.77(3).
\end{aligned}
\end{equation}
These exponents satisfy the scaling and hyperscaling relations  given by Eq.~\eqref{eq:scaling_hyper}, taking the associated uncertainties into account.

\subsection{Non-equilibrium relaxation results}
Our NER simulations employed a hybrid MCS protocol consisting of one random Tomita MCS followed by one over-relaxation Tomita MCS to accelerate the simulation dynamics. Time was subsequently measured in units of these combined steps. 
The number of independent simulation runs (replicas) ranged from 394 to 28,000, depending on the system size and temperature. To estimate the critical point and exponents, the system with \( L = 256 \) was simulated at multiple temperatures near the critical point, using 12,500 replicas at each temperature. To estimate the error in quantities, the ensembles were divided into five parts, and the bootstrap method was used for error estimation.

\begin{figure}[tbp]
\centering\includegraphics[width=\columnwidth]{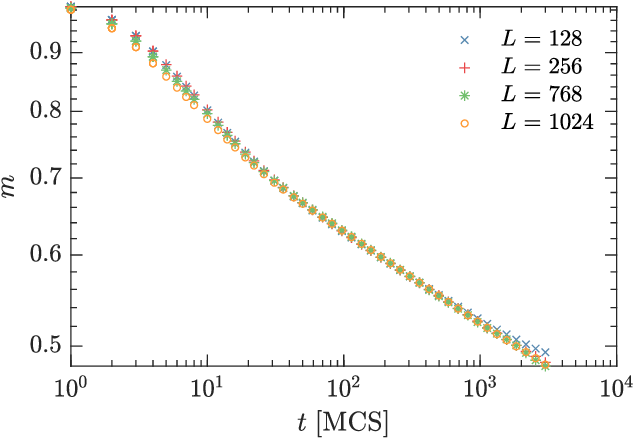}
\caption{Non-equilibrium relaxation process of magnetization for various system sizes at temperature $T=0.849$.  Only 50 data points are shown for better visualization. 
Time is measured in terms of a combined MCS, which includes one random step plus one over-relaxation step.  Error bars are omitted as they are smaller than the symbols.}
\label{fig:NER_multi_size}
\end{figure}

To determine how long the NER process remains  consistent with the infinite system limit for a given system size, we  plot the NER process for various system sizes as shown in Fig. \ref{fig:NER_multi_size}. If the NER process for a specific system size deviates from the thermodynamic limit and begins to approach equilibrium, its behavior will diverge from that of larger systems. This occurs since a larger size always follows the behavior of the thermodynamic limit for a longer time. From this figure, it can be seen that the behavior for size \( L=128 \) deviates from the larger system sizes near the end of the process.
In contrast, the NER processes for systems with \( L=256 \), \( L=768 \), and \( L=1024 \) overlap closely, indicating consistency with the thermodynamic limit.

\begin{figure}[tbp]
\centering	\includegraphics[width=\columnwidth]{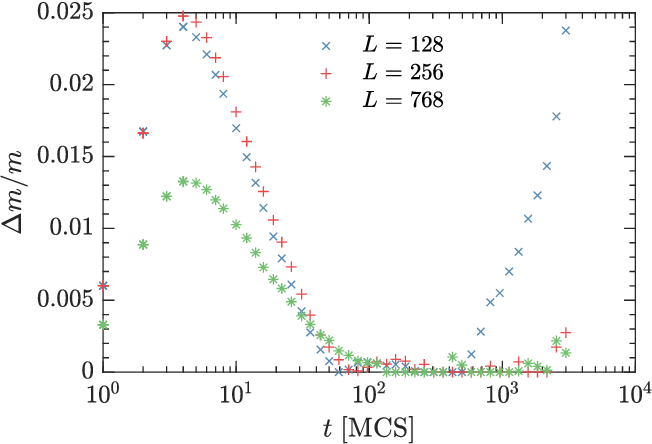}
\caption{Temporal variation of the relative difference in magnetization compared to $L=1024$ for the data in Fig. \ref{fig:NER_multi_size}. Other details are the same as in Fig. \ref{fig:NER_multi_size}.}
\label{fig:NER_multi_size_err}
\end{figure}

To precisely determine how long the NER process follows the infinite system behavior for smaller systems, we plotted the relative difference in magnetization compared to the \( L=1024 \) system in Fig. \ref{fig:NER_multi_size_err}. The difference is computed as the minimum distance while accounting for the error range, such that it equals zero when the values agree within their respective uncertainties.  As shown in the figure, the NER process for the system with $L=128$ behaves identically to that of $L=1024$ between $t=100$ and $t=500$. After this range, its behavior diverges and no longer follows the thermodynamic limit. However, systems with sizes $L=256$ and $768$ exhibit thermodynamic limit behavior up to at least $t=3000$. Additionally, Figs. \ref{fig:NER_multi_size} and \ref{fig:NER_multi_size_err} indicate that the NER processes reach the asymptotic behavior at a time around $t \simeq 100$. 

\begin{figure}[tbp]
\centering	\includegraphics[width=\columnwidth]{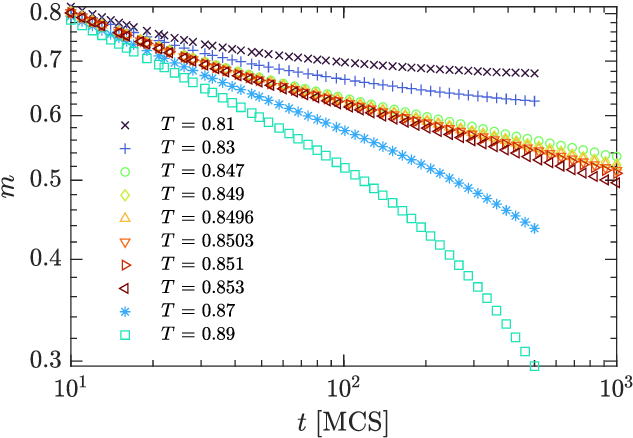}
\caption{The NER processes of magnetization at different temperatures are shown. Simulations for temperatures \( T = 0.849 \) to \( T = 0.851 \) used a system size \( L = 256 \), whereas simulations for other temperatures used \( L = 128 \). The number of plotted data points has been reduced for better visibility. Other details are the same as in Fig. \ref{fig:NER_multi_size}.}
\label{fig:NER_multi_temp}
\end{figure}

The NER processes of magnetization for multiple temperatures are plotted in Fig. \ref{fig:NER_multi_temp}. As can be seen, the curves for temperatures $T = 0.81$ and $T = 0.83$ are concave upward, indicating that these temperatures are below the critical point. On the other hand, the curves for $T = 0.87$ and $T = 0.89$ are concave downward, suggesting they are above the critical point. However, determining the concavity of the curves for temperatures near the critical point (from \( T = 0.847 \) to \( T = 0.853 \)) is not feasible through simple visual inspection of the graphs. Therefore, we employed three criteria to identify the concavity. For these temperatures, the process is examined over the interval  \( t = 100 \) to \( 1000 \) for $L=256$  to ensure asymptotic behavior has been reached. Note that in all these criteria, the data were analyzed on a full logarithmic scale.

In the first criterion, the start and end of the interval are connected by a straight line, and the number of points on the curve lying above or below the line is counted.  If most points lie above the line, the concavity is downward; if most lie below, the concavity is upward. Note that in this approach, we do not consider all points on the curve. Instead, 50 points are selected at equal intervals on the logarithmic scale because, on this scale, the density of data points increases near the endpoint. If points were not selected at equal intervals on this scale, the region near the endpoint would be overemphasized, skewing the result.

\begin{figure}[tbp]
\centering\includegraphics[width=\columnwidth]{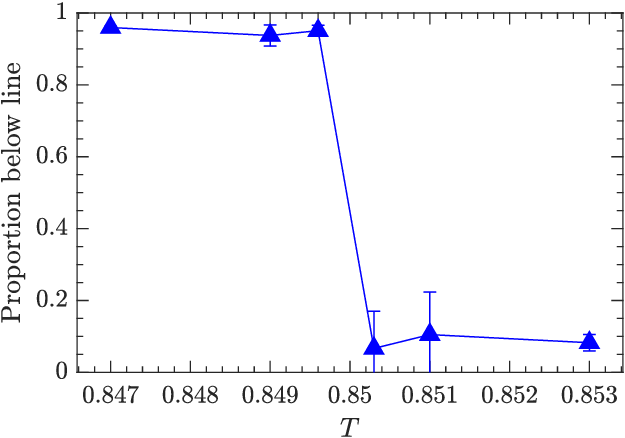}
\caption{The proportion of points below the line connecting the points at $t=100$  and $t=1000$ in the NER of magnetization, for temperatures $T=0.847$ to 0.853 in Fig.~\ref{fig:NER_multi_temp}. The analysis is performed on a log-log scale. The lines are drawn as visual guides only.}
\label{fig:NER_prop_below}
\end{figure}

The proportion of points below the line plotted against temperature is shown in Fig. \ref{fig:NER_prop_below}. It is observed that for temperatures $T=0.8496$ and lower, most points are below the line, while for temperatures $T=0.8503$ and higher, most points are above the line. As a result, this plot suggests that the critical temperature lies between $T=0.8496$ and $T=0.8503$.

\begin{figure}[tbp]
\centering\includegraphics[width=\columnwidth]{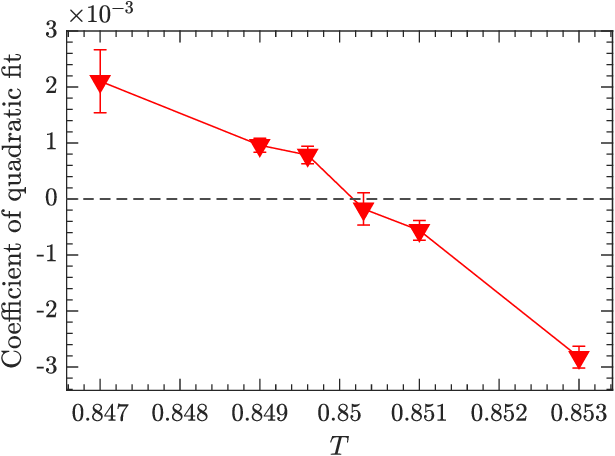}
\caption{The coefficient of quadratic fit to the NER of magnetization at temperatures $T=0.847$ to 0.853 indicated in Fig.~\ref{fig:NER_multi_temp}. The fitting is performed on a log-log scale over the time interval $t=[100, 1000]$. The  lines serve only as guides to the human eye, with the dashed horizontal line indicating a zero coefficient value.}
\label{fig:NER_quadratic}
\end{figure}

In the second criterion, similar to the first one, we selected 50 points  at equally spaced logarithmic intervals.  A second-degree polynomial was fitted to these points, and the coefficient of the quadratic term was determined. If this coefficient is positive, the curve is concave upward; if negative, it is concave downward.  The quadratic coefficient plotted as a function of temperature is shown in Fig. \ref{fig:NER_quadratic}.  The concavity is upward at temperatures \( T = 0.8496 \) and below, but downward at \( T = 0.851 \) and above. However, at \( T = 0.8503 \), the quadratic coefficient is approximately zero, and the concavity cannot be conclusively determined due to the error margin. As a result, this criterion implies that the critical point lies between  $T=0.8496$ and $0.851$.

In the third criterion, we examined the trends in the slopes of the NER curves. To ensure smooth results, each curve was divided into seven overlapping  segments, with each segment spanning one-third of the total interval. The slope of each segment was calculated by selecting 25 points at equal logarithmic intervals and determining the slope of the linear fit to these points.

\begin{figure}[tbp]
\centering\includegraphics[width=\columnwidth]{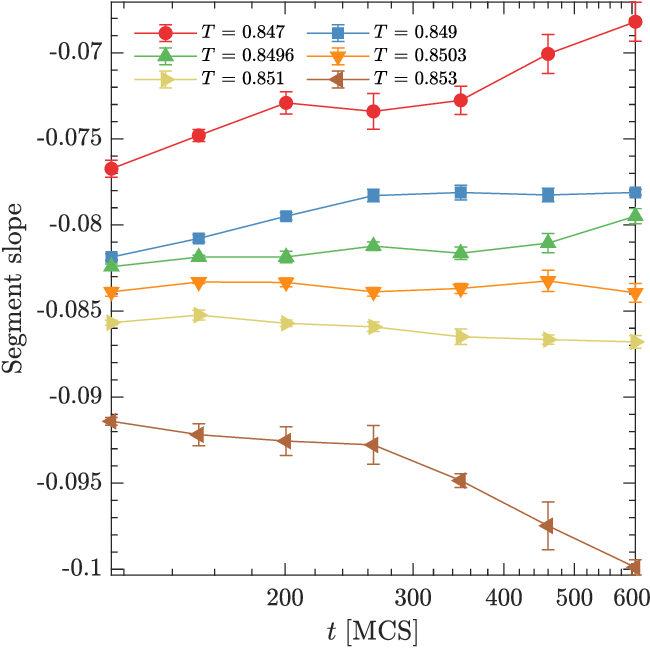}
\caption{The slopes of segments in the NER process of magnetization, plotted against the midpoints of the segments for temperatures $T=0.847$ to 0.853 shown in Fig.~\ref{fig:NER_multi_temp}. The segments overlap, with each spanning one-third of the total interval. This analysis is performed on a full logarithmic scale over the time interval  $t=[100, 1000]$. The lines are only guides to human eye.}
\label{fig:NER_slope}
\end{figure}

The plot of the slope for a segment as a function of the midpoint of the segment for different temperatures is shown in Fig. \ref{fig:NER_slope}.  For temperatures \( T = 0.8496 \) and below, the slopes of the segments exhibit an increasing trend, whereas for \( T = 0.851 \) and above, they display  a decreasing trend. At temperature \( T=0.8503 \), whether the slope trend is increasing or decreasing cannot be determined due to the error margin. From this, we infer that the critical point lies between \( T = 0.8496 \) and \( T = 0.851 \).

\begin{figure}[tbp]
\centering\includegraphics[width=\columnwidth]{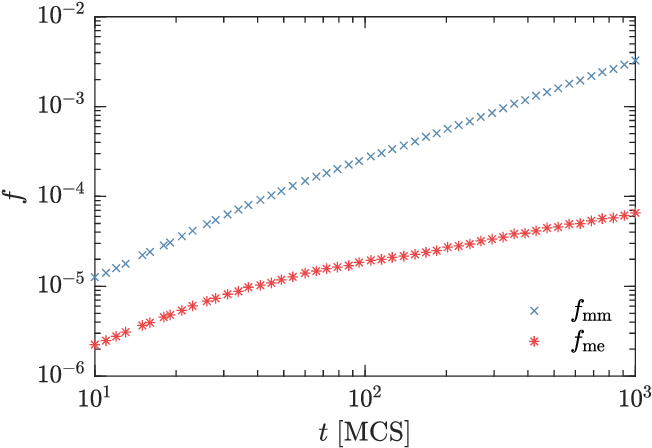}
\caption{The NER processes of \( f_\mathrm{mm}(t) \), and \( f_\mathrm{me}(t) \)  at estimated \( T_\mathrm{c} = 0.8503 \) for lattice size $L=256$  are shown. Other details are the same as in Fig. \ref{fig:NER_multi_size}.}
\label{fig:f_mm_f_me}
\end{figure}

Based on the three aforementioned criteria, we determine the critical point to be \( T_\mathrm{c}=0.8503(6) \).  To calculate the critical exponents, we used quantities \( f_\mathrm{mm}(t) \),  \( f_\mathrm{me}(t) \) (plotted in Fig. \ref{fig:f_mm_f_me}), and \( m(t) \). The slopes of these quantities are then calculated on a logarithmic scale within the time interval \( t = 100 \) to \( t = 1000 \) at \( T = 0.8503 \), using 100 data points equally spaced on the logarithmic scale. Substituting these slopes into Eqs. \eqref{eq:exponent_by_lambda}, we obtain the critical exponents:
\begin{equation}\label{key}
z = 1.81(6), \quad \nu = 0.99(4), \quad \beta = 0.15(1).
\end{equation}
The uncertainties in the exponents arise from errors in the slope calculations and the uncertainty in the critical point estimation. 

To evaluate the influence of the chosen time interval on the results, we also calculated the critical temperature and exponents using time intervals of $t=[70, 1000]$, $t=[70, 500]$, and $t=[200, 1000]$  (see Table \ref{tab:interval}).  We observed that the calculated values are statistically consistent across all intervals, though the uncertainties increase for shorter time intervals.

\begin{table}
\centering
\caption{Critical temperature \(\tc\) and exponents (\(z\), \(\nu\), \(\beta\)) computed using NER processes over different time intervals. }
\begin{tabularx}{\columnwidth}{XXXXX}
	\toprule
	$t$        & $\tc$      & $z$     & $\nu$   & $\beta$ \\  \midrule
	$[100\; 1000]$ & 0.8503(6)  & 1.81(6) & 0.99(4) & 0.15(1) \\
	$[70\; 1000] $ & 0.8503(6)  & 1.82(3) & 1.00(2) & 0.15(1) \\
	$[70\; 500]  $ & 0.8510(15) & 1.80(3) & 1.00(3) & 0.15(1) \\
	$[200\; 1000]$ & 0.8500(30) & 1.90(20)  & 1.02(4) & 0.16(2) \\ \bottomrule
\end{tabularx}
\label{tab:interval}
\end{table}

\section{Discussion and conclusion}
This study reports reliable and consistent estimates for the critical temperature and critical exponents of a triangular lattice of XY magnetic dipoles using two independent methods: the equilibrium Monte Carlo simulations with histogram reweighting and the non-equilibrium relaxation method.  With the equilibrium method, the critical parameters are estimated as
$\tc = 0.8491(4)$, $ \gamma = 1.77(3)$, $ \beta = 0.160(7)$, and $\nu = 1.05(2)$.
The non-equilibrium relaxation method yields similar results:
$\tc = 0.8503(6)$, $ \nu = 0.99(4)$, and $\beta = 0.15(1)$, along with the dynamical exponent $z = 1.81(6)$. 
These values are notably close to the 2D Ising universality class, though the value of $\beta$ exhibits a slight deviation from that of the Ising class which might indicate a new universality class (see Table \ref{tab:scaling}).

\begin{table}
\centering
\caption{ Critical parameters for the 2D triangular lattice of $O(2)$  magnetic dipoles. For some values, the error was not reported in the original reference. ``Eqm" denotes equilibrium results.}
\label{tab:scaling}
\begin{tabularx}{\linewidth}{p{2.6cm}XXXX}
	\toprule
	Reference              & $T_\mathrm{c}$ & $\nu$             & $\beta$            & $\gamma$    \\ \midrule
	Rastelli et al. \cite{rastelli2002} & $0.88$         & $1$               & 0.21(3)            & 1.70(3)     \\
	$\mathrm{Tomita}^a$  \cite{tomita2009}   & 0.6695(7)      & $1.10(2)^\dagger$ & $0.148(5)^\dagger$ & $1.93(4)^{\dagger*}$ \\
	Our results (Eqm)        & 0.8491(4)      & 1.05(2)           & 0.160(7)           & 1.77(3)     \\
	Our results (NER)       & 0.8503(6)      & 0.99(4)           & 0.15(1)            & $1.68(8)^*$ \\[2ex] 
	2D Ising \cite{goldenfeld1992}      &      & 1          & 1/8            & 7/4 \\ \bottomrule
\end{tabularx}
\flushleft 
\footnotesize \vspace{-0.5em}
\textsuperscript{a} Simulation performed for $O(3)$ dipoles. \\
$^\dagger$ Values calculated from $1/\nu$ and $\beta/\nu$  reported in the original reference. \\
$^*$ Calculated using the hyperscaling relation $\nu d = 2\beta + \gamma$.
\end{table}

Our results extend previous studies on the Triangular lattice of dipoles by using larger system sizes and more advanced analytical and simulation techniques, achieving higher precision in determining critical parameters and more accurate estimates of their uncertainties. Table \ref{tab:scaling} summarizes our findings and compares them with prior research. Notably, earlier studies relied on finite-size scaling collapse and Binder parameter analysis of multiple simulations at different temperatures, which typically offer lower accuracy than the approaches used in this study \cite{ferrenberg1988, ozeki2007} and may have even led to an underestimation of the reported uncertainties.

One notable work in this area was conducted by Rastelli et al. \cite{rastelli2002}, who investigated triangular lattice spin systems, including the purely dipolar case, using system sizes up to \( L = 36 \). Our critical parameter estimates for the dipolar triangular lattice align closely with their reported values, though their estimates have larger uncertainties due to the smaller system sizes they investigated. Additionally, they did not report uncertainty estimates for some values.

In another work, Tomita determined critical parameters for  \(O(3)\) dipoles  (3D-rotating) on various lattices, including the triangular one, using systems up to  \(L = 96\) \cite{tomita2009}. Unlike Tomita's study,  our dipoles are restricted to the lattice plane, resulting in a different critical temperature (see Table \ref{tab:scaling}). Despite this difference, the critical exponents for \(O(2)\) and \(O(3)\) dipoles are expected to be identical, as dipoles in a 2D lattice align within the lattice plane in the ground state. For the critical exponents, Tomita’s results show reasonable agreement with ours, though some discrepancies exist. Therefore,  further investigations are needed to confirm the hypothesis of universal critical behavior for $O(2)$  and $O(3)$  dipolar systems. 

In conclusion, this study employs advanced equilibrium and non-equilibrium Monte Carlo methods to investigate the XY dipolar  triangular lattice. These approaches yield precise and mutually corroborating estimates for the critical temperature and exponents, surpassing the accuracy of previous studies. Our findings contribute to a deeper understanding of phase transitions in 2D dipolar systems, with relevance to the behavior of magnetic thin films, and magnetic nanoparticle arrays.

\begin{acknowledgments}
This research received support from the IASBS Grant G2024IASBS12644. AI tools were utilized for grammar correction, sentence rephrasing, Persian-to-English translation, and overall text enhancement.
\end{acknowledgments}

%

\end{document}